\documentclass[
aps,%
12pt,%
final,%
notitlepage,%
oneside,%
onecolumn,%
nobibnotes,%
nofootinbib,% In the current version of REVTeX, when this option is on,
superscriptaddress,%
noshowpacs,%
centertags]%
{revtex4}

\def\refitem#1{\relax}

\begin{document}
\title{On the critical behavior of (2+1)-dimensional QED}

\author{\firstname{A.V.} \surname{Kotikov}}
\email{kotikov@theor.jinr.ru}
\affiliation{Bogoliubov  Laboratory of Theoretical Physics, JINR Dubna, 
141980 Dubna, Russia}

\begin{abstract}

It is shown the analysis \cite{Koti} for
QED in 2+1 dimensions with N four-component fermions
in the leading and next-to-leading
orders of the $1/N$ expansion.
As it was demonstrated in  \cite{Koti}
the range of the admissible values $N$, where the dynamical
fermion
mass exist, decreases strongly with the increasing of the gauge
charge.
So, in Landau gauge the dynamical chiral symmetry breaking appears for $N<3.78$,
that is very close to the results of the leading order and in
Feynman gauge dynamical mass is completely absent.
\end{abstract}

\maketitle

%\section{Introduction}

Quantum Electrodynamics in 2+1 dimensions (QED$_3$) has acquired increasing
attention \cite{Koti}-\cite{6} because of its similarities to (3+1)
dimensional QCD. Moreover, last years a new strong interest comes to QED$_3$
in the relation with graphene properties (see \cite{6.1} and 
discussions and references therein).
Graphene, a one-atom-thick layer of graphite, is a remarkable system with many
unusual properties that was fabricated for the first five years ago \cite{6a}.
Theoretically it was shown long time ago \cite{6b} that quasiparticle
excitations in graphene are described by the massless Dirac equation in
(2+1) dimension. This explains why the bilayer graphene in external fields
is a subject of intensive recent study \cite{6c}.

A number of
investigations have been performed for the study of dynamical chiral symmetry
breaking in QED$_3$ and very different results have been obtained. Using the
leading order (LO) in the $1/N$ expansion of the Schwinger-Dyson (SD) equation,
Appelquist et al. \cite{1} showed that the theory exhibits a critical
behavior as the number $N$ of fermion flavors approaches $N_c = 32/ \pi^2$;
that is, a fermion mass is dynamically generated only for $N<N_c$. On the
contrary, Pennington and collaborators \cite{2},
adopting a more general non-perturbative approach to the SD equations, found
that the dynamically generated fermion mass decreases
exponentially with $N$, vanishing only as $N \rightarrow \infty$. This
conclusion was supported also by Pisarski \cite{3} by the use of the other
methods.
On the other hand, an alternative non-perturbative study by
Atkinson et al. \cite{4} suggested that chiral symmetry is unbroken at
sufficiently large $N$. The theory has also been simulated on the lattice
\cite{5,6}. Remarkably, the conclusions of Ref. \cite{5} are in the agreement
with the
existence of a critical $N$ as predicted in the analysis of Ref. \cite{1}
while the second paper \cite{6} contains the opposite results.

Because the critical value $N_c$ is not large, the contribution of the higher
orders in the $1/N$ expansion can be essential and may lead to better
understanding of the problem. The purpose of this work is to consider the $1/N$
correction \cite{10,Koti} to 
%Appelquist et al. \cite{1} 
LO result \cite{1}.\\

{\bf 1.} The Lagrangian of massless QED$_3$ with $N$ flavors is
$$ L = \overline \Psi ( i \hat \partial - e \hat A ) \Psi
- \frac{1}{4} F_{ \mu \nu}^2,$$
where $ \Psi$ is taken to be a four component complex spinor.
In massless case, which we are considering,
the model contains infrared divergences, which can be canceled when the model
is analyzed in a $1/N$ expansion \cite{7}. Since the theory is massless,
the mass scale is the dimensional coupling constant $a = Ne^2/8$ which is
kept fixed as $N \rightarrow \infty$.

Following \cite{1} we study the solution of the SD equation.
The inverse fermion propagator has the form
%$$S(p)^{-1} = - \hat p [1 + A(p)] + \Sigma (p),$$
$$S^{-1}(p) = - [1 + A(p)] \left( \hat p + \Sigma (p) \right),$$
where $A(p)$ is the wave-function renormalizable coefficient and
$ \Sigma (p)$ is a dynamical, parity-conserving mass taken to be the same for
all the fermions.

The SD equation is
%has the form
\begin{eqnarray}
\Sigma (p) = \frac{2a}{N} Tr \int \frac{d^3 k}{(2 \pi )^3}
\frac{ \gamma^{\mu} D_{\mu \nu}(p-k)
%\bigl( \hat k
\left[1 + A(k) \right]
\bigl( \hat k
+ \Sigma (k) \bigr)
\Gamma^{\nu}(p,k)}{
%k^2
\left[1 + A(k) \right]^2
\left( k^2 + \Sigma^2(k)
\right)}, \label{1} \end{eqnarray}
where\footnote{Following 
%the Ref. 
\cite{10} we introduce a nonlocal
gauge-fixing term. The detailed analysis of this possibility has been given in
Ref. \cite{11}.}
$$ D_{\mu \nu}(p) = \frac{g_{\mu \nu} - (1 - \xi) p_{ \mu} p_{ \nu} / p^2}{p^2
\left[1 + \Pi (p) \right]}$$
is the photon propagator and $ \Gamma ^{ \nu}(p,k)$ is the vertex function.\\

%\section{LO}

 {\bf 2.} The LO approximations in the $1/N$ expansion are
$$ A(p) = 0, ~~ \Pi (p) = a/ \mid p \mid, ~~
%\mbox{ and } 
\Gamma^{\nu}(p,k) = \gamma^{\nu},$$
where we neglect the fermion mass in the calculation of $ \Pi (p)$. The gap
equation is
\begin{eqnarray}
\Sigma (p) = \frac{8a(2 + \xi)}{N} Tr \int \frac{d^3 k}{(2 \pi )^3}
\frac{ \Sigma (k) }{ k^2
\bigl[ (p-k)^2 + a \mid p-k \mid
\bigr]}, \label{1a} \end{eqnarray}
where we ignore the term $ \Sigma^2 (k)$ in the denominator of r.h.s. .
%, as this has been done in Ref. \cite{1}.

Following 
%Ref. 
\cite{1}, we set
\begin{eqnarray}
\Sigma (k) = (k^2)^{ \alpha} \, . \label{2} \end{eqnarray}
One can see, that for large $a$ the r.h.s. of 
%eq.
(\ref{1a}) together with
condition (\ref{2}) (and the contributions of higher orders)
% also) 
can be
calculated by the standard rules for massless diagrams of the perturbation
theory (see, for example, \cite{12}). Thus, we have for large $a$
\begin{eqnarray} 1 = \frac{2 + \xi}{  \beta L}
\label{2.1} \end{eqnarray}
with $ \beta = (- \alpha)( \alpha + 1/2)$ and $ L \equiv \pi^2 N$,
or
\begin{eqnarray}
\alpha_{ \pm} =
\bigl( -1 \pm
\left[1 - 16(2 + \xi)/L \right]^{1/2}
\bigr) /4 \, .  \label{3} \end{eqnarray}

We reproduce the solution given 
%by Appelquist et al i
n Ref.
\cite{1}. That analysis yields a critical number of fermions
$ N_c = 16(2+ \xi)/ \pi^2 \approx 1.62(2+ \xi)$ (i.e. $L_c = 16(2+ \xi)$),
such that for $N>N_c ~~~ \Sigma(p) = 0$ and
$$ \Sigma(0) \simeq exp
\bigl[ -2 \pi / (N/N_c - 1)^{1/2} \bigr]$$
for $N<N_c$. Thus, chiral-symmetry breaks
%breaking occurs 
when $ \alpha$ becomes
complex, that is for $N<N_c$.

%\section{NLO}

{\bf 3.} The next-to-leading order (NLO) approximation has been included 
in \cite{Kiev,Koti}
%We calculate exactly the contributions of the NLO Feynman integrals
%(see Fig.1 from Ref. \cite{10}) 
using the differential equation method \cite{DEM}.
%rules for the
%calculation
%of massless diagrams. 
The results have a cumbersome form \cite{Kiev}, which is similar to
results for complicated massless diagrams obtained using Gegenbauer 
polynomials \cite{Gegenbauer}.
% (they contain two-
%and three-sum terms) and will be given in the separate publication. Here 
In \cite{Koti} we have 
analyzed 
%only 
simplified form, which contains only the
terms $ \sim (- \alpha)^{-k} \mbox{ and } \sim ( \alpha + 1/2)^{-k}~~~
(k=1,2,3)$ from the series  given in \cite{Kiev}. These terms are most 
important in the neighborhood of the critical point $N_c$. 
%We get the following equation
The eq. (\ref{2.1}) is replaced now by
\begin{eqnarray}
1 = \frac{(2 + \xi)}{ \beta L} +
\bigl[f( \xi) + \beta \varphi( \xi) \bigr]
\frac{1}{( \beta L)^2}, \label{5} \end{eqnarray}
where $f( \xi) = 4(1 - \xi)/3 - \xi^2,~~
\varphi ( \xi) = 176/9 - 4 \pi^2 - (16/3) \xi +4 \xi^2$.

Let us get the exact critical value $N_c$ from eq. (\ref{5}). Supposing
$ \alpha = \alpha_c \equiv -1/4$ we obtain the critical values in the
following form
\begin{eqnarray}
N_{c, \pm} = \frac{8}{ \pi^2}
\bigl[  (2 + \xi) \pm
\left( (2 + \xi)^2 + 4f( \xi) + \varphi( \xi)/4   \right)^{1/2}
\bigr], \label{6} \end{eqnarray}
i.e.
$$N_{c,+}( \xi) = (3.31, 3.35, 3.09, 2.81) ,~~
N_{c,-}( \xi) = (-0.07, 0.38, 1.29, 1.88)$$
for
$\xi = (0.0, 0.3, 0.7, 0.9),$
respectively.

Notice the intriguing fact that follows from (\ref{6}).
The addition of $1/N$ correction leads to the occurrence of the second
critical point (for $0.05 \leq \xi \leq 0.95$) such that for
$N<N_{c,-}$ the chiral symmetry does not break. The dynamical mass
generation exists in the interval between the critical points $N_{c,-}$
and $N_{c,+}$. For $ \xi \geq 0.95$ this interval disappears and the
chiral symmetry breaking is absent.  For small values of gauge parameter
$ \xi$ ($ \xi \leq 0.05$) new critical point does not occur.
%We note also that the use of the exact solution of SD equation in the
%first two orders can change only weakly the numerical values of critical
%points $N_{c,+}$ and $N_{c,-}$ but not qualitative picture of the decrease of
%the interval, where the dynamical mass generates, with the increase of the
%value of $ \xi $.

The solution of the eq.(\ref{5}) is
$$ \beta_{ \pm} = \frac{1}{2L} ~
\Bigl[ 2 + \xi + \frac{ \varphi ( \xi)}{L} \pm
\bigl( (2 + \xi)^2 + 4f( \xi) + 2(2 + \xi) \frac{ \varphi ( \xi)}{L} +
\frac{ \varphi ^2( \xi)}{L^2}
\bigr)^{1/2}  \Bigr]$$
has the simple form in Landau gauge
\begin{eqnarray}
 \beta_{ \pm} ( \xi = 0) = \frac{1}{L} ~
\Bigl[ 1 + \frac{ \varphi (0)}{2L} \pm
\surd \overline { 7/3 }
\left( 1 + \frac{3}{14} \frac{ \varphi (0)}{L} \right)
\bigl( 1 +
\frac{ \frac{3}{49} \varphi ^2(0)/L^2}{
\left( 1 + \frac{3}{14} \frac{ \varphi (0)}{L} \right)^2}
\bigr)^{1/2}  \Bigr], \label{7} \end{eqnarray}
where the last term in r.h.s. of eq.(\ref{7}) is very small for $L \sim L_c$.
Leaving it out we get the following equation for $ \beta_+$
 $$ \beta_{+} ( \xi = 0) \approx
 1 + \surd \overline {7/3} \frac{1}{L} +
\left(1 + \surd \overline {3/7} \varphi (0)
\right) \frac{1}{2L^2}
\approx \frac{2.52}{L} \left( 1 - \frac{6.52}{L} \right),$$
which has coefficients are close to those from the paper \cite{10}.\\

%\section{Resume}
%$Resume$. 

{\bf Resume.} We reviewed the results of \cite{Koti}, where the
$O(1/N^2)$ terms have been included 
into SD equation 
%exactly 
and the strong gauge dependence of the result has been  found.
%which in general has not the form of eq.(\ref{4})
% (see eq. (\ref{7})). 
Hence, the addition of $1/N$ correction does not
lead to the essential improvement in the understanding of dynamical chiral
symmetry breaking. However, as it was shown in \cite{Koti},
%we note that 
in the Landau gauge
%, where Ward identities can be satisfied
%in the case of free vertex (see \cite{13}), 
the inclusion of $O(1/N^2)$ terms slightly changes only quantitative (but not
qualitative) properties of the LO results.
Thus, in the Landau gauge the analysis in \cite{Koti} gives further evidence 
in favor of the solution has been given
by Appelquist et al. \cite{1}.\\
%Hence, it seems that
%the $1/N$ expansion of the kernel of SD equation describes reliably the
%critical behaviour of the theory.\\

{\bf Acknowledgments.} Author thanks the Organizing Committee of the 
International Conference
``Critical Point and Onset of Deconfinement (CPOD)''
for invitation.


\newpage

\begin{thebibliography}{99}
\bibitem{Koti} 
\refitem{article}
A.V. Kotikov, JETP\ Lett.\ {\bf 58}, 734 (1993).
%%CITATION = ZFPRA,58,785;%%


\bibitem{1} 
\refitem{article}
T. Appelquist, D. Nash and L.C.R. Wijewardhana, Phys.\ Rev.\ Lett.\ 
{\bf 60}, 2575 (1988).

\bibitem{2} 
\refitem{article}
M.R. Pennington and D. Walsh,  Phys.\ Lett.\ B {\bf 253}, 246
 (1991);  D.C. Curtis, M.R. Pennington and D. Walsh, Phys.\ Lett.\ B
{\bf 295}, 313 (1992).

\bibitem{3} 
\refitem{article}
R. Pisarski,
 Phys.\ Rev.\ D {\bf 29}, 2423 (1984); {\bf 44}, 1866 (1991).

\bibitem{4} 
\refitem{article}
D. Atkinson, P.W. Johnson and P. Maris, Phys.\ Rev.\ D {\bf 42}, 602 (1990).

\bibitem{5} 
\refitem{article}
E. Dagotto, A. Kocic and J.B. Kogut,
Phys.\ Rev.\ Lett.\ {\bf 62}, 1083 (1989); Nucl.\ Phys.\ B {\bf 334}, 279 
(1990).

\bibitem{6} 
\refitem{article}
V. Azcoiti and X.Q. Luo, Nucl.\ Phys.\ Proc.\ Suppl {\bf 30}, 741 (1993);
V. Azcoiti, V. Laliena and X.Q. Luo, Nucl.\ Phys.\ Proc.\ Suppl {\bf 47}, 
565 (1996).

\bibitem{6.1} 
\refitem{article}
O.V. Gamayun, E.V. Gorbar and  V.P. Gusynin, 
Phys.\ Rev.\  B {\bf 81}, 075429 (2010).

\bibitem{6a} 
\refitem{article}
A.H. Castro Neto {\it et al.},
%, F. Guinea, N.M.R. Peres, K.S. Novoselov, and A.K. Geim,
Rev.\ Mod.\ Phys.\ {\bf 81}, 109 (2009);
K.S. Novoselov {\it et al.},
%, A.K. Geim, S.V. Morozov, D. Jiang, M.I. Katsnelson, 
%I.V. Grigorieva, S.V. Dubonos, and A.A. Firsov, 
Nature {\bf 438}, 197 (2005).
%University of Zaragosa, Report No. DFTUZ/92/18, 1992 (unpublished).

\bibitem{6b} 
\refitem{article}
G.W. Semenoff, Phys.\ Rev.\ Lett. {\bf 53}, 2449 (1984);
P.R. Wallace, Phys.\ Rev.\ {\bf 71}, 622 (1947).

\bibitem{6c} 
\refitem{article}
E. McCann and V.I. Fal'ko, Phys.\ Rev.\ Lett. {\bf 96}, 086805 (2006);
B.E. Feldman {\it et al.},
%, J. Martin, and A. Yacoby, 
Nature\ Phys. {\bf 5}, 889 (2009);
Y. Zhao {\it et al.},
%, P. Cadden-Zimansky, Z.Jiang, and P. Kim, 
Phys.\ Rev.\ Lett.\ {\bf 104}, 066801 (2010);
E.V. Gorbar {\it et al.},
%, V.P. Gusynin, and V.A. Miransky, 
JETP\ Lett.\ {\bf 91}, 314 (2010).

\bibitem{10} 
\refitem{article}
D. Nash, Phys.\ Rev.\ Lett. {\bf 62}, 3024 (1989).


\bibitem{7} 
\refitem{article}
T. Appelquist and R. Pisarski, Phys.\ Rev.\ D {\bf 23}, 2305 (1981);
%\bibitem{8} 
%\refitem{article}
R. Jackiw and S. Templeton, Phys.\ Rev.\ D {\bf23}, 2291 (1981);
T. Appelquist and U. Heinz, Phys.\ Rev.\ D {\bf 24}, 2169 (1981).

%\bibitem{9} 
%\refitem{article}
%T. Appelquist, M. Bowick, D. Karabali and
% L.C.R. Wijewardhana, Phys.\ Rev.\ D {\bf 33}, 3774 (1986).


\bibitem{11} 
\refitem{article}
D.V. Shirkov, Nucl.\ Phys.\ B {\bf 332}, 425 (1990).

\bibitem{12} 
\refitem{article}
D.I. Kazakov, Phys.\ Lett.\ B {\bf 133}, 406 (1983).


\bibitem{Kiev}
\refitem{article}
A.V. Kotikov, Preprint ITP-91-29E, Kiev, 1991.

\bibitem{DEM}
\refitem{article}
A.V. Kotikov, Phys.\ Lett.\  B {\bf 254}, 158 (1991);
%%CITATION = PHLTA,B254,158;%%
Phys.\ Lett.\  B {\bf 259}, 314 (1991);
%%CITATION = PHLTA,B259,314;%%
Phys.\ Lett.\  B {\bf 267}, 123 (1991).
%%CITATION = PHLTA,B267,123;%%

\bibitem{Gegenbauer}
\refitem{article}
K.G. Chetyrkin {\it et al.},
%, A.L. Kataev, F.V. Tkachov, (Moscow, INR) . 1980.
Nucl.\ Phys.\ B {\bf 174}, 345 (1980);
A.V. Kotikov, Phys.\ Lett.\  B {\bf 375}, 240 (1996).
 %%CITATION = PHLTA,B375,240;%%


%\bibitem{13} 
%\refitem{article}
%K. Kondo and H. Nakatuni, Mod.\ Phys.\ Lett. A {\bf 4}, 2155
%(1989); Nucl.\ Phys.\ B {\bf 351}, 236 (1991).

\end{thebibliography}
\end{document}